\documentclass[journal]{IEEEtran}
\usepackage{amsmath,amsfonts}
\usepackage{algorithmic}
\usepackage{algorithm}
\usepackage{array}
\usepackage[caption=false,font=normalsize,labelfont=sf,textfont=sf]{subfig}
\usepackage{textcomp}
\usepackage{stfloats}
\usepackage{url}
\usepackage{verbatim}
\usepackage{graphicx}
\usepackage{cite}
\usepackage{hyperref}
\usepackage{xcolor}
\usepackage{array} 
\usepackage{booktabs} 
\usepackage{multirow} 

\usepackage{xcolor}

\hypersetup{
    colorlinks=false,         
    citebordercolor=green,    
    pdfborder={1 1 1}         
}

\usepackage{amssymb}
\makeatletter

\newcommand{\Rmnum}[1]{\expandafter\@slowromancap\romannumeral #1@}
\makeatother

\usepackage{diagbox} 
\usepackage{tikz} 
\usepackage{makecell}
\usepackage{float}
\usepackage{tabularx}

\usepackage{threeparttable}
\usepackage{placeins} 
\usepackage{afterpage}
\usepackage{tcolorbox}

\begin{document}

\title{Generative AI-Enhanced Multi-Modal Semantic Communication in Internet of Vehicles: System Design and Methodologies}

\author{Jiayi Lu, Wanting Yang, Zehui Xiong,~\IEEEmembership{Senior Member,~IEEE}, Chengwen Xing,~\IEEEmembership{Member,~IEEE}, \\Rahim Tafazolli,~\IEEEmembership{Senior Member,~IEEE}, Tony Q.S. Quek,~\IEEEmembership{Fellow,~IEEE}, and Mérouane Debbah,~\IEEEmembership{Fellow,~IEEE}
\thanks{Jiayi Lu is with the School of Information
and Electronics, Beijing Institute of Technology, Beijing, China, and also with the Pillar of Information Systems Technology and Design, Singapore University of Technology and Design, Singapore (e-mail: lujiayi.ee@gmail.com).}
\thanks{Wanting Yang, Zehui Xiong, Tony Q. S. Quek are with the Pillar of Information Systems Technology and Design, Singapore University of Technology and Design, Singapore (e-mail: wanting\_yang@sutd.edu.sg; zehui\_xiong@sutd.edu.sg; tonyquek@sutd.edu.sg).}
\thanks{Chengwen Xing is with the School of Information
and Electronics, Beijing Institute of Technology, Beijing, China (e-mail: xingchengwen@gmail.com).}
\thanks{Rahim Tafazolli is with the Institute for Communication Systems (ICS), 5/6GIC, The University of Surrey, UK (e-mail: r.tafazolli@surrey.ac.uk).}
\thanks{M. Debbah is with KU 6G Research Center,
Khalifa University of Science and Technology, P O Box 127788, AbuDhabi, UAE (e-mail: merouane.debbah@ku.ac.ae) and also with CentraleSupelec, University Paris-Saclay, 91192 Gif-sur-Yvette, France (e-mail: merouane.debbah@ku.ac.ae).}}



\maketitle

\begin{abstract}
Vehicle-to-everything (V2X) communication supports numerous tasks, from driving safety to entertainment services. To achieve a holistic view, vehicles are typically equipped with multiple sensors. However, processing large volumes of multi-modal data increases transmission load, while the dynamic nature of vehicular networks adds to transmission instability. To address these challenges, we propose a novel framework, Generative Artificial Intelligence (GAI)-enhanced multi-modal semantic communication (SemCom), referred to as G-MSC, designed to handle various vehicular network tasks by employing suitable analog or digital transmission. GAI presents a promising opportunity to transform the SemCom framework by significantly enhancing semantic encoding, semantic information transmission and semantic decoding. It optimizes multi-modal information fusion at the transmitter, enhances channel robustness during transmission, and mitigates noise interference at the receiver. To validate the effectiveness of the G-MSC framework, we conduct a case study showcasing its performance in vehicular communication networks for predictive tasks. The experimental results show that the design achieves reliable and efficient communication in V2X networks. In the end, we present future research directions on G-MSC.
\end{abstract}


\section{Introduction}
\IEEEPARstart{W}{ith} the rapid development of the Internet of Things (IoT) and connected devices, vehicular networks have emerged as a key component of transportation systems. The Internet of Vehicles (IoV) system comprises vehicles, roadside units (RSUs), vehicular edge computing (VEC), and vehicular cloud computing (VCC). Collaboration among these entities supports the execution of various safety and entertainment tasks, such as path planning and downloading high-definition videos within the vehicle, most of which are highly dependent on communication. Different tasks may require varying communication ranges. Since a single sensor provides a limited perspective and may fail to capture information due to occlusion, it is necessary to integrate multi-view data from various sensors, such as cameras, radar, and others. Unfortunately, the substantial volume of multi-modal data, coupled with the growing number of vehicles, exacerbates the challenge of radio resource scarcity. Furthermore, the high mobility and dynamic topology of IoV nodes lead to frequent handovers, posing significant difficulties for access management\cite{xu2024integration}. Therefore, it underscores the growing importance of ensuring both efficiency and reliability in communication between users and servers in IoV systems.
\begin{figure}[t]
    \centering
    \includegraphics[scale=0.46]{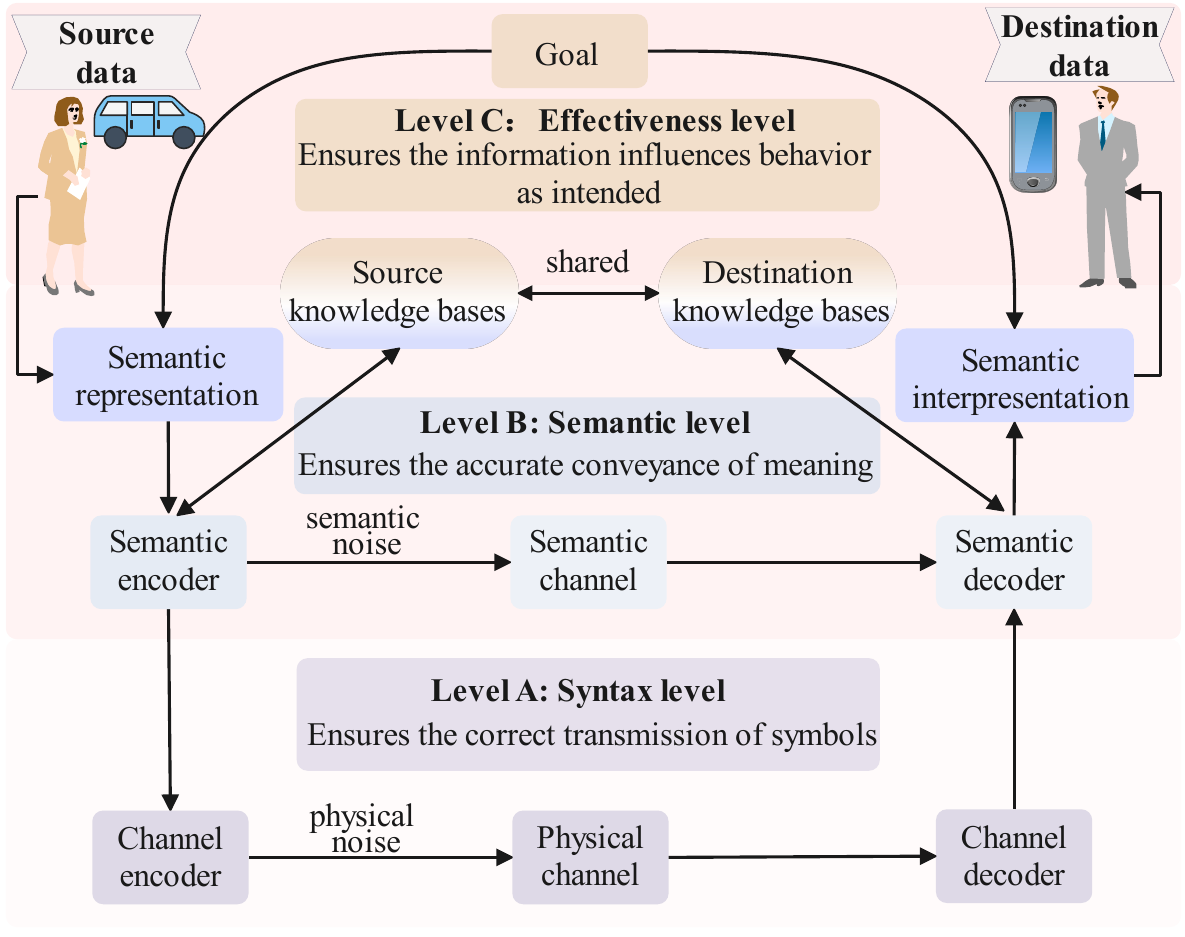}
    \caption{The three levels of information theory communication. Compared to semantic-level communication, which focuses on accurately conveying comprehensive semantic information embedded in source data, effectiveness-level communication, only focuses on transmitting the task-specific semantic information required by different vehicles in IoV.}
    \label{The three levels of information theory communication}
\end{figure}

To reduce the volume of transmitted data while preserving the accuracy of transmitted semantics, semantic communication (SemCom) has emerged as a key technology for the upcoming 6G era\cite{yang2022semantic}. The significance of semantics was first recognized by Shannon and Weaver, who introduced three levels of communication\cite{shannon1948mathematical}. As shown in Fig.\ref {The three levels of information theory communication}, traditional Level A communication focuses primarily on the accurate transmission of bits, while Levels B and C extend beyond bit-level accuracy to ensure the intended meaning is correctly conveyed. In vehicular networks, task-oriented communication drives SemCom to primarily focus on Level C.

With the development of autonomous driving, as shown in Table \Rmnum{1}, the computational power equipped in vehicles has been continuously increasing, as SemCom relies heavily on computational resources. Nonetheless, current research on SemCom predominantly focuses on traditional data types such as text, images, and videos, with relatively few SemCom architectures specifically designed for vehicular networks. \cite{xu2023semantic} introduced a multi-vehicle collaborative semantic perception framework tailored for vehicle re-identification, where nearby RSUs transmit captured images to the cloud center using SemCom to reduce transmission overhead and improve image integration efficiency. Similarly, \cite{feng2024semantic} only outlined a SemCom-enhanced edge intelligence framework for multi-vehicle, multi-modal data, but fails to offer concrete details or implementation specifics. Both studies overlook the challenges of processing data from multiple sensors in vehicles and the necessity of employing different communication manners tailored to varying computational requirements and transmission distances for different tasks. Additionally, current research on SemCom mostly adopts the classic joint source-channel coding (JSCC) framework with end-to-end training. While JSCC has proven effective in optimizing overall system performance and enhancing data transmission efficiency, its inherent nature still presents some limitations. Specifically, JSCC limits the interpretability and manipulation of semantic information. Furthermore, many tasks in vehicular networks rely on multi-vehicle collaboration, but the end-to-end training approach of JSCC in current SemCom paradigms limits the knowledge base to empirical datasets, which require costly data collection and labor-intensive labeling for multi-user scenarios\cite{yang2024rethinking}.

Thankfully, the flourishing Generative Artificial Intelligence (GAI)\hypersetup{pdfborder={0 0 0}}\footnote{We summarize the typical GAI technologies and their characteristics in Table $\textnormal{\Rmnum{2}}$.}\hypersetup{pdfborder={1 1 1}} presents a transformative opportunity to reinvent the SemCom framework\cite{zhang2024generative}. Unlike traditional JSCC approaches, where background knowledge merely supports the training of semantic encoders and decoders, the generative SemCom paradigm harnesses background information to play a crucial role in semantic representation and inference, ensuring more accurate preservation of original semantic content at the receiver. In addition, GAI plays a pivotal role in facilitating digital transmission by optimizing data compression and reconstruction, making it compatible with modern digital communication systems. Moreover, GAI can enhance channel robustness by strengthening channel modeling and increasing channel estimation accuracy to ensure reliable communication in the dynamic environment of vehicular networks. 

\begin{table}[t]
\caption{Autonomous driving computational power requirements.}
\centering
\renewcommand{\arraystretch}{1.5} 
\setlength{\arrayrulewidth}{0.4pt} 
\setlength{\tabcolsep}{10pt} 
\begin{threeparttable}
\begin{minipage}[t]{0.48\textwidth}
\centering
\resizebox{\textwidth}{!}{ 
\begin{tabular}{|c|c|c|}
\hline
\textbf{Level} & \textbf{Function} & \textbf{TOPS} \\ \hline
L0 & \makecell[c]{\textit{No Automation}\\ Human-driven} & \diagbox{}{} \\ \hline
L1 & \makecell[c]{\textit{Driver Assistance Automation}\\ Human-driven\\ Lane departure and adaptive cruise control \\ cannot be implemented simultaneously} & \textless 5 \\ \hline
L2 & \makecell[c]{\textit{Partial Automation}\\ Human-driven\\ Lane departure and adaptive cruise control \\ can be implemented simultaneously} & \textless 10 \\ \hline
L3 & \makecell[c]{\textit{Conditional Automation}\\ Human can refrain from driving \\ but must take over when requested} & 30 \raisebox{-0.5ex}{\textasciitilde} 60 \\ \hline
L4 & \makecell[c]{\textit{High Automation}\\ The vehicle can operate without a driver \\ but is limited to specific road and environment} & \textgreater 100 \\ \hline
L5 & \makecell[c]{\textit{Full Automation}\\ The vehicle can operate without a driver \\ and is not restricted to specific road} & \textgreater 1000 \\ 
\end{tabular}
}
\end{minipage}\hfill
\begin{minipage}[t]{0.48\textwidth}
\centering
\resizebox{\textwidth}{!}{ 
\begin{tabular}{|c|c|c|c|}
\hline
\textbf{Autonomous driving}& \textbf{GPU} & \textbf{TOPS} & \textbf{Achieved Level} \\ \hline
NIO et7 & 4 $\cdot$ Orin-X & 1016 & L2 \raisebox{-0.5ex}{\textasciitilde} L3 \\ \hline
Li Auto L9 & 2 $\cdot$ Orin-X & 508 & L2 \raisebox{-0.5ex}{\textasciitilde} L3 \\ \hline
XPENG G9 & 2 $\cdot$ Orin-X & 508 & L2 \raisebox{-0.5ex}{\textasciitilde} L3 \\ \hline
AITO M9 &1 $\cdot$ MDC610 & $200^{\ast}$ & L2 \raisebox{-0.5ex}{\textasciitilde} L3 \\ \hline
IM Motors L6 &1 $\cdot$ Orin-X & 254 & L2 \raisebox{-0.5ex}{\textasciitilde} L3 \\ \hline
PPA 01 & 2 $\cdot$ Orin-X & 508 & L2 \raisebox{-0.5ex}{\textasciitilde} L3 \\ \hline
Apollo GO & unannounced & 1200 & L4 \\ \hline
\end{tabular}
}
\end{minipage}

\begin{minipage}[t]{0.47\textwidth}
\vspace{0.2cm} 
\footnotesize 
\hspace{0.2cm} 
$^{\ast}$ It utilizes dense computing power, which is equivalent to twice the sparse computing power. Others refer to sparse computing power. Although the computational power of existing autonomous vehicles supports Level 4 and Level 5 autonomous driving, algorithms to fully realize these levels of automation have yet to be developed.
\end{minipage}

\end{threeparttable}
\end{table}
To bridge the research gap in GAI-enhanced multi-modal SemCom in IoV, we embark on this study. In this article, we propose a framework called GAI-enhanced multi-modal SemCom, abbreviated as G-MSC. The framework is designed to support the transmission of multi-modal data and handle a wide range of tasks within the IoV. The primary contributions of our works are as follows:
\begin{itemize}
\item[$\bullet$] We outline the four types of vehicle-to-everything (V2X) communication for various tasks. Then, we explore the typical multi-modal data types used in autonomous driving and discuss key challenges.
\item[$\bullet$] We detail the proposed G-MSC architecture for analog or digital communication transmission according to the specific requirements of different tasks in IoV. For various tasks, we explain how GAI enhances semantic encoding, channel, and semantic decoding. Additionally, we provide an overview of various methods of multi-modal data fusion, comparing their respective pros and cons.
\item[$\bullet$] We conduct a case study on a simplified analog transmission framework within the G-MSC for predictive tasks in vehicular networks. We apply Bird's Eye View (BEV) fusion to process multi-modal data in the semantic encoder and employ diffusion model (DM) for refinement in the semantic decoder. We also implement a predictive task, generating forecasted BEV images. To evaluate the performance, we adopt Intersection over Union (IoU) as the evaluation metric. Experimental results show that diffusion-based images significantly improve IoU and visual clarity, further validating its effectiveness.
\end{itemize}

The remainder of this article is organized as follows. In Section \Rmnum{2}, we introduce various tasks and data modalities in autonomous driving and discuss the key challenges. In Section \Rmnum{3}, we detail the proposed G-MSC architecture. Then, in Section \Rmnum{4} we present a case study on a simplified analog transmission framework within the G-MSC. Finally, we conclude the article and highlight potential future research directions in Section \Rmnum{5}.

\section{Communication-Dependent Tasks in IoV }
In this section, we first highlight four types of V2X communication. Next, we introduce various modalities of data involved in communication in IoV system. Then, we summarize the key challenges in terms of semantic encoder, transmission, and semantic decoder at the end of the section.

\begin{table*}[t]
\caption{Summary of the typical GAI technologies.}
\centering
\renewcommand{\arraystretch}{1.5} 
\setlength{\arrayrulewidth}{0.5pt} 
\setlength{\tabcolsep}{10pt} 
\begin{threeparttable}
\LARGE
\resizebox{\textwidth}{!}{ 


\begin{tabular}{|c|c|c|}
\hline
\textbf{Technologies} & \textbf{Architectures} & \textbf{Features}  \\ \hline \makecell[c]
{\textbf{Generative Adversarial}\\ \textbf{Network (GAN)}} & \makecell[l]{{\textbullet} \textit{Generator Network}: Generate data that closely resembles real data.\\{\textbullet} \textit{Discriminator Network}: Evaluate the authenticity of the generated \\data and engages in adversarial training with the generator.\\ $\ast$  {\textbf{Variants}}: \textit{Conditional GAN}, \textit{Wasserstein GAN}, \textit{StyleGAN}} & \makecell[l]{Generate high-quality images,\\but training is unstable.}  \\ \hline
\makecell[c]{\textbf{Variational}\\ \textbf{Autoencoder (VAE)}}& \makecell[l]{{\textbullet} \textit{Encoder Network}: Map input to a latent space distribution.\\{\textbullet} \textit{Decoder Network}: Reconstruct the data from samples in the latent \\space to approximate the original input.\\ $\ast$  {\textbf{Variants}}: \textit{Vector Quantized VAE}, \textit{Variational Auto-Encoding GAN}} & \makecell[l]{Stabilize training, \\but generate images with \\lower sample quality.} \\ \hline
\makecell[c]{\textbf{Diffusion Model}\\ \textbf{(DM)}}& \makecell[l]{{\textbullet} \textit{Forward Process}: Gradually add noise to the original data in a \\ step-by-step process.\\ {\textbullet} \textit{Denoising Process}: Revert the noise from the normal distribution. \\ $\ast$  {\textbf{Variants}}: \textit{Stable Diffusion}, \textit{Denoising Diffusion Probabilistic Model}} & \makecell[l]{Generate high-quality images, \\with stable training.}  \\ \hline
\makecell[c]{\textbf{Transformer-based}\\ \textbf{Model}}& \makecell[l]{{\textbullet} \textit{Self-Attention}: Learns the dependencies between elements in the \\ input by computing attention weights. \\ $\ast$  {\textbf{Variants}}: \textit{Generative Pre-trained Transformer (GPT)}} & \makecell[l]{Capable of handling long-range \\ dependencies parallelizable training.}  \\ \hline
\end{tabular}
}

\end{threeparttable}
\end{table*}

\subsection{Typical Communication Tasks}
\begin{figure*}[t]
    \centering
    \includegraphics[scale=0.36]{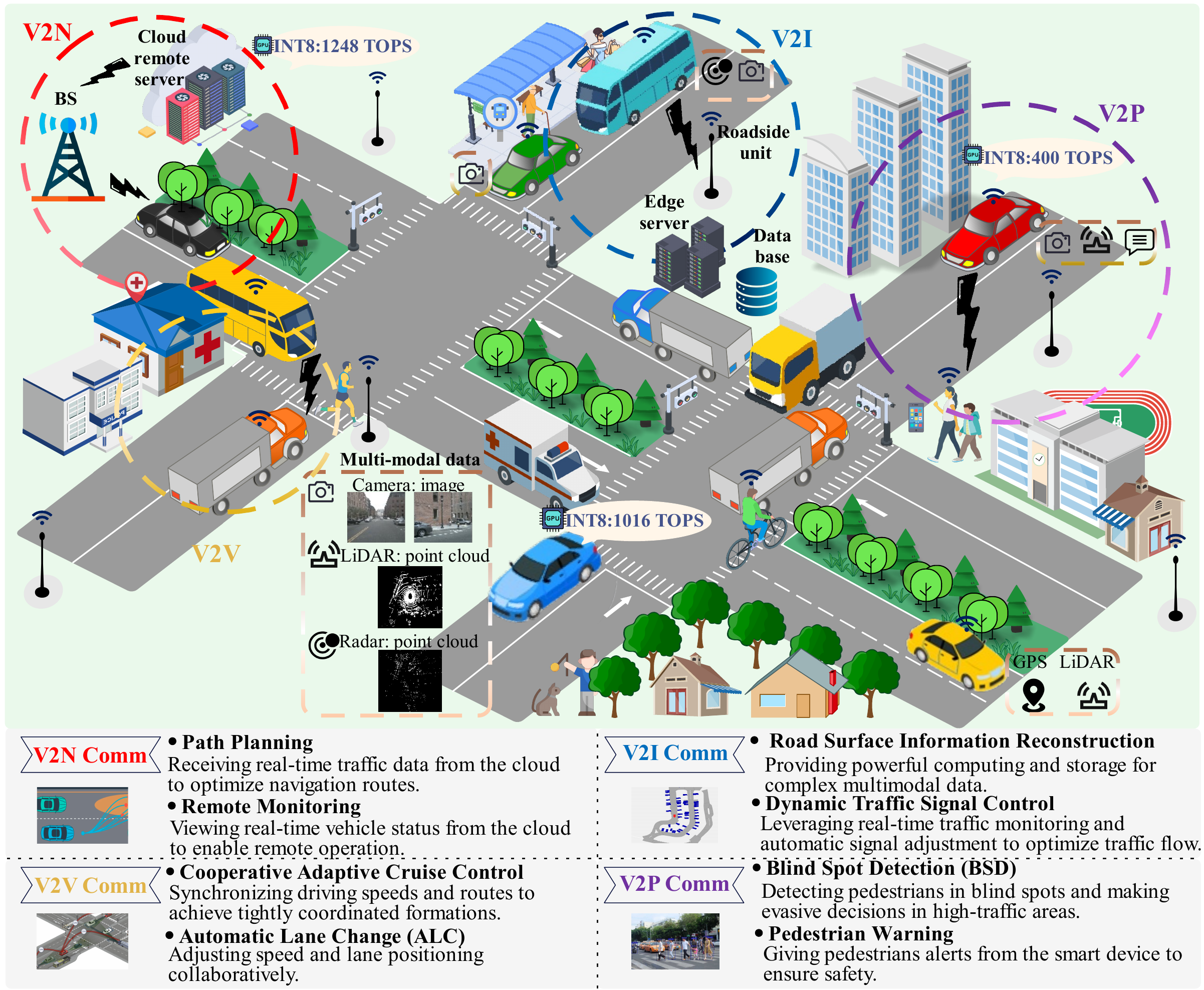}
    \label{fig_first_case}
    \caption{A realistic Internet of Vehicles scenario with multi-modal V2X communication.}
    \label{Multi-Modal Multi-Task Communication in Realistic Internet of Vehicles Scenarios} 
\end{figure*}

In vehicular networks, safety-related tasks, collaborative offloading tasks, and entertainment tasks rely on efficient communication. For example, in densely populated areas, vehicles may struggle to detect all pedestrians due to obstructed views, making communication with pedestrians' smart devices crucial for safety. When a vehicle’s computational capacity is exceeded by extensive multi-modal data, offloading processing to the edge server via RSU becomes essential. Additionally, when autonomous vehicles need to change lanes, more detailed road information is required. In these situations, communication between vehicles is crucial for gathering necessary data and ensuring safety. 
Based on different types of receivers, V2X communication can be categorized into  the following four types, as shown in Fig.\ref{Multi-Modal Multi-Task Communication in Realistic Internet of Vehicles Scenarios}:
\begin{itemize}
\item[$\bullet$] \emph{Vehicle-to-network (V2N) communication:} In vehicular networks, VCC is suited for tasks that require real-time data retrieval from cloud servers, such as remote monitoring, path planning, and high-definition video downloads within the vehicle. These tasks necessitate high bandwidth, long-range communication, multi-hop forwarding, and substantial computational power. Under the circumstances, end-to-end communication based on JSCC is prone to signal attenuation and noise interference, which can degrade data recovery quality. Therefore, digital transmission is commonly preferred in this scenario.
\item[$\bullet$] \emph{Vehicle-to-infrastructure (V2I) communication:} VEC is employed for low-latency communication scenarios, supporting real-time decision-making. For example, Tasks such as road surface information reconstruction, which need processing large volumes of multi-modal data, typically require substantial computing power. To overcome the computational limitations of some vehicles, part of the processing is offloaded to nearby edge servers. Vehicle sensors collect traffic data and transmit it to edge computing devices for integration and processing. The primary challenge in V2I communication is ensuring low latency and reliable data transfer with limited bandwidth.
\item[$\bullet$] \emph{Vehicle-to-vehicle (V2V) communication:} Autonomous driving tasks such as Cooperative Adaptive Cruise Control (CACC) and Automatic Lane Change (ALC) require vehicle-to-vehicle information exchange to collaboratively adjust speed and lane positioning, ensuring safety during lane changes. In addition to CACC and ALC, V2V communication is essential for tasks like platooning, where vehicles move in coordination to reduce fuel consumption and improve traffic efficiency. These tasks depend on real-time sharing of data, such as speed and braking status, enabling rapid responses to environmental changes. Communication typically uses Dedicated Short-Range Communication. However, due to high vehicle speeds, significant channel variations or signal blockages from occlusions may occur in V2V communication.
\item[$\bullet$] \emph{Vehicle-to-pedestrian (V2P) communication:} In high pedestrian traffic zones, such as areas near schools and hospitals, the Blind Spot Detection (BSD) task plays a critical role. Analog transmission is well-suited for such short-range tasks due to its high fidelity and cost-effectiveness. By obtaining pedestrian location data from smartphones and other smart devices, vehicles can better detect pedestrians or objects in blind spots, enabling timely braking and yielding. 
\end{itemize}

\subsection{Typical Communication Data Modalities}
In the communication tasks discussed earlier, multiple data modalities, such as images, point clouds, text, and speech are often involved to ensure information completeness and transmission reliability. Specifically, different modalities of data exhibit distinct characteristics and limitations. Camera sensors provide detailed visual information but lack depth perception and are susceptible to weather conditions. Radar sensors are effective in various weather and road conditions but offer lower resolution. LiDAR sensors deliver precise depth information, though at a higher cost. Inertial measurement unit (IMU) sensors detect motion changes rapidly but may drift over time, leading to potential errors. Given the unique strengths of each sensor, combining multiple sensors can maximize their benefits while reducing redundancy from overlapping information across different modalities. Moreover, passengers can control the vehicle using voice commands or by inputting text for navigation and traffic inquiries. These modalities can integrate with sensor data from radar, cameras, and others, providing a more comprehensive view of the vehicle while facilitating enhanced environmental perception and more intelligent decision-making.

\begin{table}[t]
\caption{Large-scale publicly autonomous driving datasets.}
\centering
\renewcommand{\arraystretch}{1.5} 
\setlength{\arrayrulewidth}{0.5pt} 
\setlength{\tabcolsep}{3.5pt} 
\begin{threeparttable}
\centering
\begin{tabular}{|c|c|c|} 
\hline
\textbf{Dataset} & \textbf{Sensor Configuration} & \textbf{Research Focus} \\ \hline
KITTI & \makecell[c]{4 Cameras, 1 LiDAR, 1 IMU} & \makecell[c]{Object detection,\\ visual tasks.} \\ \hline
OPV2V & \makecell[c]{4 Cameras, 1 LiDAR, 1 IMU} & \makecell[c]{Complex scenarios\\within V2V tasks.} \\ \hline
nuScenes & \makecell[c]{6 Cameras, 1 LiDAR, 5 Radars} & \makecell[c]{High-quality data\\ across a wide range of\\ driving scenarios.} \\ \hline
Lyft Level 5 & \makecell[c]{7 Cameras, 3 LiDARs} & \makecell[c]{3D perception,\\ and path planning\\ in complex \\urban environment.} \\ \hline
Waymo & \makecell[c]{5 Cameras, 5 LiDARs} & \makecell[c]{Long-range \\object tracking,\\ and behavior prediction.} \\ \hline
ApolloScape & \makecell[c]{6 Cameras, 2 LiDARs, 1 IMU} & \makecell[c]{Scene parsing,\\ self localization \\and inpainting.} \\ \hline
MapillaryVistas & \makecell[c]{Various camera sensors} & \makecell[c]{Semantic segmentation.} \\ \hline
\end{tabular}
\end{threeparttable}
\end{table}

Having reviewed the strengths and limitations of various sensors and data modalities, we study autonomous driving utilizing integrated datasets. Large-scale publicly available autonomous driving datasets, each designed for specific research purposes as outlined in Table \Rmnum{3}, are valuable for simulating diverse driving scenarios.

\subsection{Challenges and Solutions}
To effectively address the complexities of multi-modal data processing and other difficulties mentioned above in V2X communication, we introduce GAI as a solution. In the following, we outline the challenges, focusing on the transmitter, channel, and receiver.
\begin{itemize}
\item[$\bullet$] \emph{Semantic encoding for multi-modal data processing:} The diverse data in autonomous driving provides valuable information but also imposes significant communication burdens. Additionally, overlapping content across multi-modal data is common, which further complicates the processing. The variety of data types across different modalities requires varying processing times, potentially leading to delays. Therefore, it is essential to efficiently manage multi-modal data through proper alignment and fusion.
\begin{figure*}[t]
    \centering
    \includegraphics[scale=0.40]{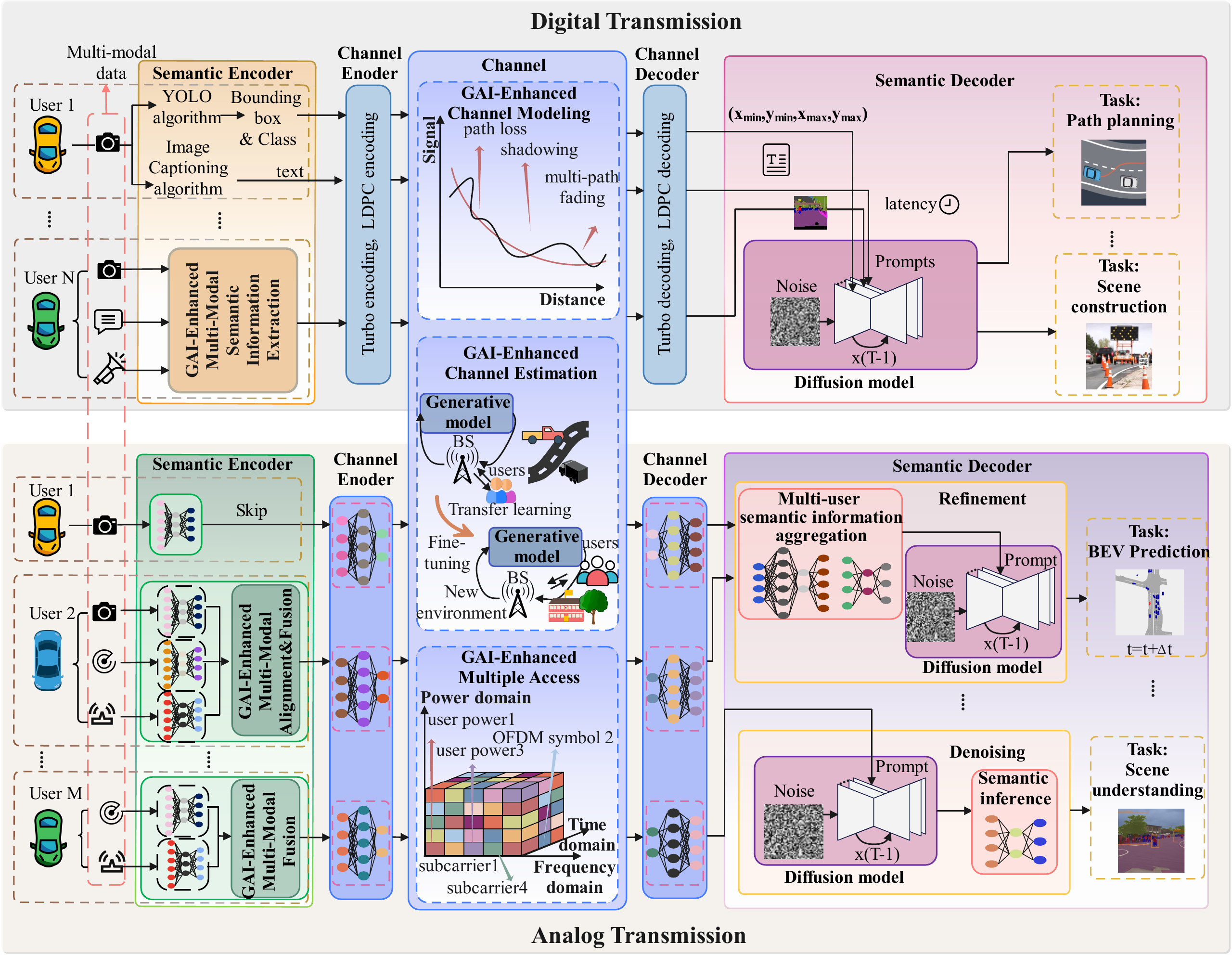}
    \label{fig_second_case}
    \caption{The proposed G-MSC framework for cooperative multi-vehicle, multi-modal SemCom via analog or digital transmission according to specific tasks.}
    \label{GAI enhanced hybrid digital-analog transmission} 
\end{figure*}
\item[$\bullet$] \emph{Channel instability:} V2X enables a variety of downstream tasks as mentioned above. Unfortunately, the high mobility and dynamic topology of the IoV system introduce unpredictable changes in the network, rendering traditional channels unsuitable for this scenario. Therefore, there is a need to develop a robust channel capable of adapting to rapidly moving vehicles.
\item[$\bullet$] \emph{Semantic decoding for noisy data:} Buildings and adverse weather conditions can cause significant signal attenuation and noise, degrading the quality of received data. Using effective methods for denoising received data is an urgent issue that needs to be addressed.
\end{itemize}

To further tackle the aforementioned problems, we employ  SemCom to reduce data transmission and detail the specific methods used for the GAI-enhanced semantic encoder, channel, and semantic decoder in the following section.

\section{Framework Design of G-MSC}
In this section, we present an overview of our proposed G-MSC framework, which consists of three key components: GAI-enhanced semantic encoder, GAI-enhanced transmission, and GAI-enhanced semantic decoder. The framework is divided into three parts to address the key challenges of processing multi-modal data, high-mobility environments, and noisy data. Additionally, to accommodate varying communication ranges and data modalities, selecting the appropriate transmission method is crucial. For long-range communication, such as V2N communication, digital transmission is preferred for its data security benefits. Conversely, in short-range and high-fidelity scenarios, analog transmission may be more effective as it avoids the leveling-off and cliff-edge effects shown in digital transmission. We explore the differences between encoders and decoders in these two transmission modes, as shown in Fig. \ref{GAI enhanced hybrid digital-analog transmission}.

\subsection{GAI-Powered Multi-Modal Semantic Encoder}
\subsubsection{Analog encoding}
The primary approaches for processing multi-modal data in vehicular networks encompass multi-modal alignment and multi-modal fusion. Multi-modal alignment involves converting various modalities spatially into a unified coordinate system and temporally synchronizing them to a common timestamp. In vehicular networks, since data from different sensor modalities are collected simultaneously by the same vehicle, temporal alignment is typically inherent. Spatial alignment methods mainly involve using rotation and translation to map data from one modality to the coordinate system of another or use the camera's intrinsic and extrinsic parameters to achieve spatial alignment between images and point clouds. Multi-modal alignment is commonly utilized as an essential step before fusion.

Multi-modal fusion is typically categorized into early fusion and late fusion approaches. Early fusion techniques that project radar data into the camera's domain may have significant distortions, whereas projecting camera data into the radar point cloud domain may affect semantic density. These early fusion methods also face challenges related to data robustness, for example, when a vehicle navigates uneven terrain, variations in radar or camera extrinsic may lead to misalignment between point clouds and images. Additionally, late fusion processes each modality independently before integration, which can lead to drawbacks such as larger model sizes, increased computational demands, and suboptimal information utilization. 

\begin{figure}[t]
    \centering
    \includegraphics[scale=0.66]{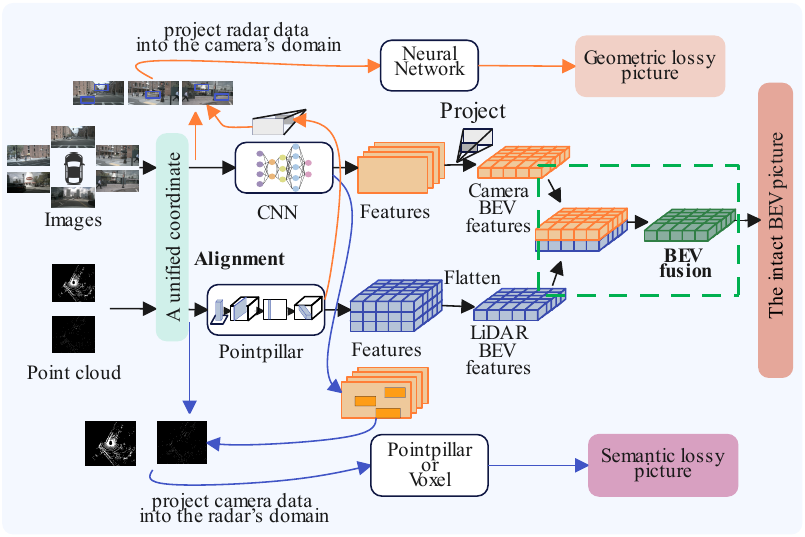}
    \caption{Multi-modal fusion methods in vehicular networks. The top orange line is for projecting radar point cloud data to camera dimensions for fusion, the bottom blue line is for projecting camera image data to radar point cloud dimensions for fusion, and the middle is for separate processing and BEV fusion of both modalities.}
    \label{Multi-modal fusion moethod}
\end{figure}

To address these issues, an intermediate fusion approach, such as BEV Fusion has been proposed for vehicular networks\cite{liang2022bevfusion}. This technique processes camera and radar point cloud data separately, using convolutional neural networks to extract features from images and employing voxel or pointpillar methods for point clouds before projecting these features into a unified BEV space to fusion, as depicted in Fig.\ref{Multi-modal fusion moethod}. Common fusion methods include addition, averaging, concatenation, ensemble, and mixture of experts\cite{feng2020deep}. The unified representation helps alleviate the distortions associated with early fusion and mitigates the substantial overheads of late fusion. Additionally, generative models like Variational Autoencoder (VAE) can be used to learn a shared latent space, mapping data from different modalities into a common coordinate space before fusion, further enhancing the effectiveness of multi-modal integration.

\subsubsection{Digital encoding}
Long-range communication typically relies on digital communication due to its strong noise resistance and high fidelity. For multi-modal data, such as voice commands and text inputs from passengers in autonomous vehicles, and image data from base stations, generative models can serve as digital encoders to generate unified textual representations. For instance, Generative Adversarial Network (GAN) like Attentional Generative Adversarial Networks (AttnGAN) \cite{xu2018attngan} can convert image data into textual descriptions. Similarly, OpenAI Whisper can transcribe speech into text, facilitating seamless integration with existing textual information. If only single-modality information is available such as image data, the multi-modal fusion can be skipped. Techniques such as the YOLO algorithm can covert image data into bounding boxes and object classes for digital transmission directly, and the image captioning algorithm can generate textual descriptions from images, particularly for tasks like autonomous navigation and scene construction. After digital encoding, the transmitted information encompasses text, bounding boxes, and classes.

\subsection{GAI-Powered High-Mobility Wireless Transmission}
Channel coding is vital for improving the reliability of communication transmission. In digital transmission, commonly used channel encoding and decoding methods include Turbo coding, Low Density Parity Check (LDPC) coding, and Huffman coding. In analog transmission, JSCC techniques have gained attention, where the channel encoder and decoder are jointly optimized through end-to-end deep learning, enabling efficient data transmission and reconstruction. Beyond channel coding, the following three aspects can further enhance performance with the assistance of GAI.
\subsubsection{GAI-enhanced channel modeling}
Traditional channel modeling typically employs additive white Gaussian noise (AWGN) and Rayleigh fading channels, which may not accurately reflect the complexities of real-world channel conditions. The authors in\cite{smith2019communication} proposed a GAN-based method for channel density estimation to handle complex channels with interference and non-linear distortions. The GAN is trained without any prior knowledge of the channel, while the generator creates approximations of the channel, and the discriminator differentiates between real and generated channels. This process allows the generator to learn realistic channel characteristics. Accurate channel modeling is vital for optimizing transmission, enabling systems to adapt to varying conditions, reduce error rates, and improve data throughput.
\subsubsection{GAI-enhanced channel estimation}
In vehicular networks with real-time channel variations, traditional estimation methods often rely on pilot-based channel estimation, which requires prior knowledge of the channel distribution. However, these methods face limitations such as increased overhead and sensitivity to noise. Moreover, the dynamic nature of channel state information necessitates frequent recalculations, potentially impacting real-time performance in communication systems. To address this problem, the DM can be employed\cite{zhang2024decision}. The model can automatically learn the mapping between received pilot signals and channel matrices in a model-free manner, enabling real-time channel restoration and significantly improving channel estimation efficiency under different distribution conditions. Accurate channel estimation is vital for reliable communication, particularly in the dynamic and unpredictable environments of vehicular networks.
\subsubsection{GAI-enhanced multiple access}
To improve spectral efficiency and address the issue of scarce radio resources, GAI can enhance various access algorithms. \cite{ma2024semantic} proposed a discrete semantic feature division multiple access paradigm using deep learning to extract multi-user semantic information into discrete subspaces, and designed a multi-user semantic interference network for inference and image reconstruction tasks. If the framework integrates GAI, it enables adaptive optimization of semantic feature extraction and enhanced semantic compression, providing an efficient solution for bandwidth utilization, particularly with growing user density and quality of service demands.

\subsection{GAI-Powered Multi-Task Semantic Decoder}
\subsubsection{Analog decoding}
There are two approaches to incorporating DM in the analog decoder. The first is denoising before semantic decoding, which serves a similar function to channel equalization in traditional communication and can be implemented as a sub-module of semantic decoders. This module learns the distribution of the channel input signals and utilizes the acquired knowledge to eliminate channel noise. The second approach is refinement, which improves the quality of decoded data. In this case, perceptual loss is often used to correct semantic errors and enhance coherence. Both methods provide flexibility and optimize the decoding process, catering to varying needs and conditions in vehicular communication systems.
\begin{figure*}[t]
    \centering
    \includegraphics[scale=0.50]{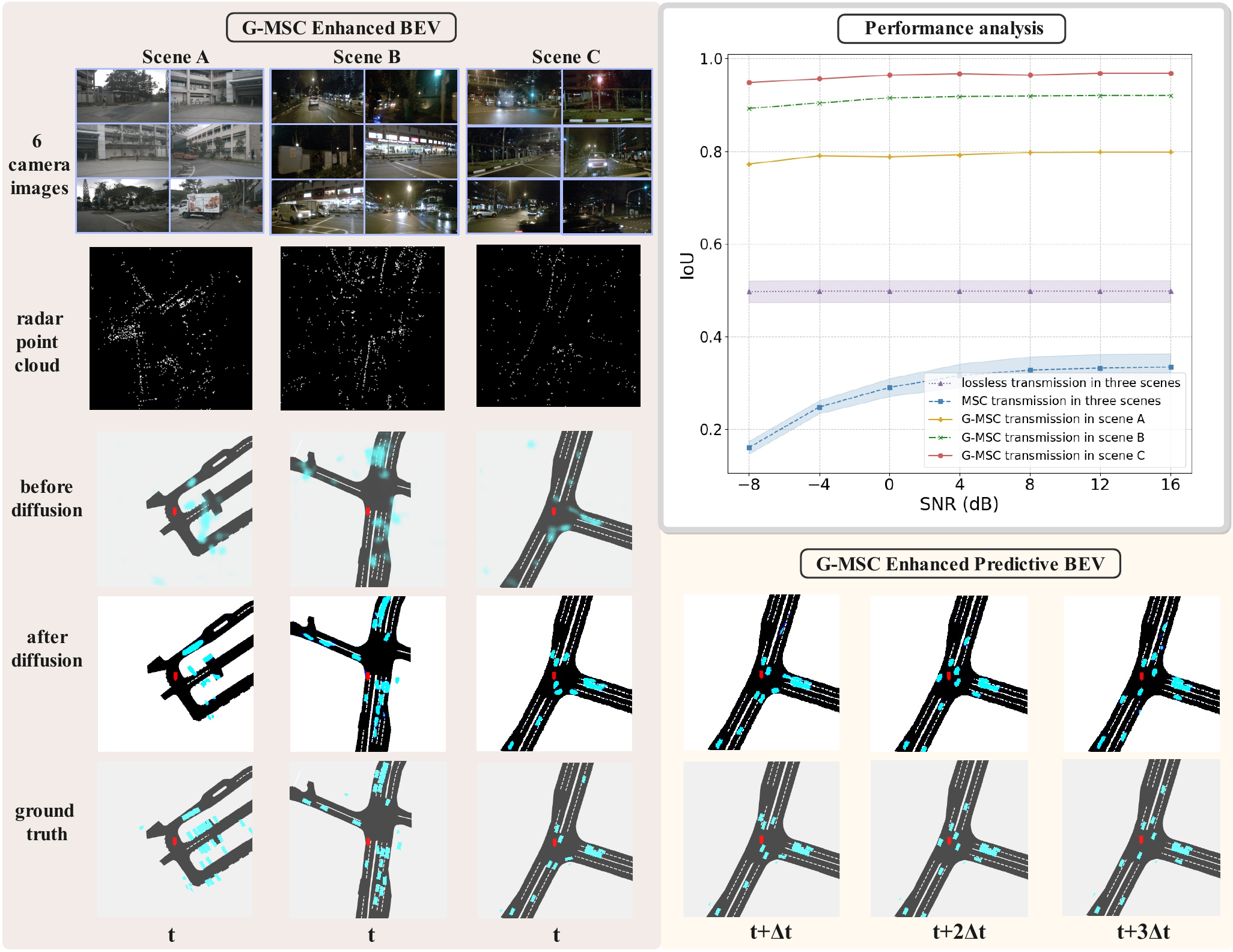}
    \caption{Experimental results of diffusion model-enhanced BEV fusion for image generation and prediction tasks. The lines for lossless channel transmission and MSC transmission without DM are represented with mean and variance, as the IoU values across the three scenarios show little difference. To emphasize the impact of DM on performance and facilitate comparison, separate lines are plotted for each of the three scenarios after applying DM.} Additionally, $\triangle t$, $2\triangle t$, and $3\triangle t$ represent future time frames, with $\triangle t$ denotes one second.
    \label{result}
\end{figure*}
\subsubsection{Digital decoding}
In digital transmission, DM can serve as an independent semantic decoder, leveraging its ability to model complex distributions and perform sequential inference. This enables DMs to handle user latency variations in collaborative tasks by processing available data immediately and integrating delayed inputs as they arrive. For example, in multi-user collaborative scene construction, vehicles capture data from various sensors. DMs can immediately process available data and integrate delayed inputs as they arrive, enabling real-time scene reconstruction. This ensures efficient and flexible collaboration in dynamic vehicular environments. Additionally, other GAI technologies, such as VAE can also enhance semantic decoding to effectively denoise, providing a stable training process to recover original data.

\section{Case Study}
In this section, we focus on a simple instance of the G-MSC framework for the predictive BEV task in analog transmission. On the transmitter side, we employ BEV fusion to integrate point cloud data and image data into a unified representation after feature extraction. On the receiver side, we utilize a DM to refine and enhance the image quality.

\subsection{Experiment Setups}
We use nuScenes-mini dataset\cite{caesar2020nuscenes}, which contains 6 cameras, 5 radar sensors, and 1 LiDAR sensor, covering 10 scenes from 4 countries, each lasting 20 seconds. The dataset contains approximately 40 images per scene, resulting in 404 images. We select 162 images from 6 scenes for training and 242 images from 4 scenes for testing.

In our experiment, we use one radar sensor and six camera sensors to perform V2I safety-related tasks. Firstly, the radar point cloud data and camera image data are transformed into the same coordinate space. Next, we voxel the point cloud data, and extract image features using a ResNet101 encoder. The 3D feature network is obtained through bilinear sampling. Afterward, the point cloud and image data are projected into the same BEV space, concatenated, and processed using channel encoding, which consists of four convolutional layers, each with five kernels. The encoded data is passed through an AWGN channel and compressed into a BEV feature map. The AWGN channel is used because we assume that power control and equalization can compensate for fading with accurate channel modeling, approximating a Gaussian channel\cite{yang2024diffusion}. The feature map is then decoded using a ResNet18 network. Finally, the resulting image is refined by DM, which enhances visual quality by removing blurring. Beyond refinement, the DM also facilitates prediction by generating not only images for the current time frame but also forecasting BEV images for the next 1, 2, and 3 seconds, which is achieved by inputting the corresponding timestamps of three consecutive future frames as prompts during training. Although the DM introduces some generation delay, this is acceptable in scenarios such as cloud centers aggregating and analyzing all information, in-car entertainment systems, and road condition prediction.

The training of the proposed system is divided into three stages. In the first stage, we train the multi-modal fusion encoder and semantic decoder modules. In the second stage, we add channel and channel encoding, and channel decoding training before the BEV compressor. The network before the channel simulates local processing on the vehicle, while the network following the channel, including the BEV compressor, simulates processing on an edge server, which handles higher computational demands. This split-learning approach addresses the challenge of limited computational power in autonomous vehicles. The learning rate is set to $1 \times 10^{-4}$, with the AdamW optimizer employed with a learning rate scheduler. In the third stage, the DM is used for refinement. The learning rate is adjusted to $1 \times 10^{-3}$, the inference is set to 1000 steps, and the Adam optimizer is adopted. Our experiments are conducted on a system equipped with an NVIDIA RTX 4090(24GB) GPU.

\subsection{Numerical Results and Analysis}
The experiment results are presented in Fig.\ref{result}. We provide visual comparisons across three different scenarios, showing images before and after diffusion-based refinement alongside the ground truth. Scene A is a daytime setting at a crowded bus station and parking lot. Scene B captures a nighttime scene with a high-speed vehicle on a wide street, and Scene C features a complexly difficult lighting scenario, where a vehicle is preparing to make a turn while observing oncoming traffic. The generated BEV image before diffusion-based refinement exhibits significant blurring, making it difficult to accurately detect vehicle positions. After diffusion-based BEV, the clarity of the images is substantially improved.

Next, we present the IoU results under different signal-to-noise ratio (SNR) conditions, comparing the images using three transmission schemes: lossless transmission without DM, MSC transmission without DM, and G-MSC transmission. IoU is a crucial metric for evaluating BEV vehicle semantic segmentation and quantitatively assesses the enhancement effect of DM by calculating the intersection-over-union between the experimental results and the ground truth. The results show that due to channel attenuation, the average IoU of images after transmission decreases across the three scenes, with IoU values improving as SNR increases. Comparing IoU curves without DM, refinement with DM shows significant improvement across all scenes, especially at low SNR, mitigating performance loss from poor channel quality. This demonstrates the strong noise resistance and generative capabilities of GAI. Furthermore, the IoU values in scene C, which has lower complexity than scenes A and B, are slightly higher after DM. Despite the higher vehicle density in scenes A and B, DM still significantly improves IoU compared to non-diffusion-processed images, underscoring its effectiveness.

Additionally, to fill generation delays and improve driving safety, we conduct the prediction task that enables vehicles to proactively adjust their behavior, avoiding collisions and optimizing routes. We use scene C to forecast images for the next 1, 2, and 3 seconds. The results show that the predicted images are clear and well-defined.

\section{Future Directions}
In this section, we provide additional insights by outlining three key future directions for GAI-enhanced SemCom in IoV.
\subsection {Hybrid Digital-Analog Transmission}
In cases where high-fidelity scenario communication uses analog transmission, transmitting short and urgent packets in this manner would waste bandwidth and time. Instead, these packets can be sent via digital communication, underscoring the need to explore hybrid analog-digital SemCom in IoV. The combined use of analog and digital transmission methods employs analog transmission for high-fidelity primary information and digital transmission for short, urgent packets as auxiliary transmission. GAI can enhance semantic information extraction at the transmitter or denoise at the receiver.

\subsection {Multi-Vehicle Semantic Informartion Scheduling}
In vehicular networks, the high density of vehicles may cause data congestion when multiple vehicles attempt to transmit simultaneously. Combining GAI with reinforcement learning can optimize the order and priority of information transmission by dynamically adjusting the scheduling of semantic information, reducing data conflicts, and improving transmission efficiency. GAI enhances the system's ability to predict complex data patterns, while reinforcement learning adapts to real-time network changes, ensuring continuous optimization, resource efficiency, and reduced latency.
\subsection{Cross-Task Coordination}
We discuss various tasks in Section \Rmnum{2}, such as pedestrian safety, and in-vehicle entertainment. Current research mainly focuses on single task-oriented communication. However, coordinating multiple tasks may be beneficial in real-world scenarios. For example, obstacle detection and lane change assistance both require similar environmental data, allowing tasks to share semantic information and improve efficiency. Investigating how GAI can enhance cross-task communication by dynamically adjusting resource allocation is essential.

\section{Conclusions}
In this article, we proposed a novel G-MSC framework for various downstream tasks in IoV, incorporating GAI-enhanced semantic encoder with multi-modal integration and semantic decoder. Moreover, we explore GAI-enhanced channel modeling, reduced channel estimation overhead, and improved multi-access performance. Depending on the specific task requirements in vehicular networks, we employed either analog or digital transmission. The framework leverages GAI technology to reduce data volume from diverse modalities in IoV and enhance communication reliability in dynamic, high-mobility vehicular networks. It also mitigates the unavoidable noise. The effectiveness of G-MSC was demonstrated through a case study on predictive BEV tasks in analog transmission, in which the BEV fusion is used to combine point cloud data and image data, and the DM is applied to enhance image quality. Results from both image clarity and IoU calculation metrics showed that GAI effectively enhanced BEV images and achieved accurate predictions. Finally, considering the outstanding performance of GAI-enhanced SemCom in IoV, we highlighted promising future directions.



\newpage

 





\end{document}